\documentclass[aps,twocolumn,superscriptaddress,showpacs,prl]{revtex4}
\usepackage{graphicx,amsfonts,amsmath,color,amsbsy,amssymb}

\begin{document}
\title{Delayed self-regulation leads to novel states in epigenetic landscapes}

\author{Mithun K. Mitra}
\email{mithun@bose.res.in}
\affiliation{S. N. Bose National Centre for Basic Sciences, Salt Lake, Kolkata 700098, INDIA.}

\author{Paul R. Taylor}
\affiliation{Systems Biology Doctoral Training Centre, Oxford University, Oxford, OX1 3QU, United Kingdom.}

\author{Chris J. Hutchison}
\affiliation{School of Biological and Biomedical Sciences, Durham University, Durham, DH1 3LE, United Kingdom.}

\author{T. C. B. McLeish}
\affiliation{Department of Physics and Astronomy, Durham University, Durham, DH1 3LE, United Kingdom.}

\author{Buddhapriya Chakrabarti}
\email{buddhapriya.chakrabarti@durham.ac.uk}
\affiliation{Department of Mathematical Sciences, Durham University, Durham, DH1 3LE, United Kingdom.}

\date{\today}
\begin{abstract}
The epigenetic pathway of a cell as it differentiates from a stem cell state to a mature lineage-committed one has been historically understood in terms of Waddington's landscape, consisting of hills and valleys. The smooth top and valley-strewn bottom of the hill represents their undifferentiated and differentiated states respectively. Although mathematical ideas rooted in nonlinear dynamics and bifurcation theory have been used to quantify this picture, the importance of time delays arising from multistep chemical reactions or cellular shape transformations have been ignored so far. We argue that this feature is crucial in understanding cell differentiation and explore the role of time delay in a model of a single gene regulatory circuit. We show that the interplay of time-dependant drive and delay introduces a new regime where the system shows sustained oscillations between the two admissible steady states. We interpret these results in the light of recent perplexing experiments on inducing the pluripotent state in mouse somatic cells. We also comment on how such an oscillatory state can provide a framework for understanding more general feedback circuits in cell development.
\end{abstract}


\maketitle

\section{Introduction}

The ``biological impossibility'' of reprogramming adult somatic cells to the pluripotent state had been accepted as a dogma for a long time in biology~\cite{Solter:84}. This view was radically changed by the work of John B. Gurdon in 1962 who showed that a nucleus from a fully differentiated frog intestinal epithelial cell could generate a functioning tadpole upon transplantation into an enucleated egg~\cite{Gurdon:62,Gurdon:06}. In another seminal work, Shinya Yamanaka and co-workers demonstated for the first time in 2006, that four transcription factors (Sox4, Oct2, Klf-4 and c-Myc) were capable of reprogramming an adult mouse fibroblast cell to pluripotency ~\cite{Yamanaka:06}. These induced pluripotent stem cells (iPSC) were fully germline-competent and were used to clone fully functioning adult mice ~\cite{Yamanaka:07, Jaenisch:07, Wernig:07}. The discovery of germline-competent iPSCs has opened up a new avenue for understanding the process of cellular differentiation, besides offering a new source for developing stem cells for tissue regeneration and other biomedical applications, without the ethical concerns of harvesting embryonic stem cells. Transcription factor based somatic cell reprogramming has since been shown to be a robust process, and human pluripotent cells have also been developed from somatic cells using a combination of transcription factors, using the SOKM protocol ~\cite{Yamanaka:07} as well as using other TFs such as NANOG and Lin28 in place of Klf-4 and c-Myc ~\cite{Yu:07, Thomson:09}. While induced pluripotency has been characterised for a number of different cell lines, understanding the key gene regulatory networks and molecular mechanisms that underlie the process remains a key outstanding challenge ~\cite{Nagy:10,Rizzino:10,Hochedlinger:10}.

Cell development and differentiation has been interpreted in light of Waddington's epigenetic landscape~\cite{Waddington:57}, visualized as a set of marbles rolling down a hill with the position of the marble indicative of the state of cellular development. Thus undifferentiated cells all start at the same state at the top of the hill and end up in different valleys corresponding to their differentiated states at the bottom of the hill depending on the surface topography. These differentiated cell states are separated by barriers which prohibit their spontaneous transformation from one state to another. Though visually compelling and despite past attempts a quantification of Waddington's landscape has been attempted only recently~\cite{Wang:11,Ferrell:12}.

Cell developmental circuits have been modeled as self-regulatory networks, where a transcription factor promotes its own production~\cite{Ferrell:12,Wang:11} as well as inhibits the production of other TFs (in multi-variable models) ~\cite{Wang:11}. Such TF regulated gene networks are known to accurately represent cell fate decision pathways in biological models. A two variable self-activating and mutually inhibiting gene network, has been found in various tissues where a multipotent cell undergoes a binary decision process~\cite{Huang:10,Enver:07,Wang:11}. One known instance is when the Common Myleoic Progenitor (CMP) differentiates into either the myeloid or the erythroid fates, depending on the expression levels of the PU.1 and the GATA1 transcription factors~\cite{Enver:09,Enver:07,Wang:11}. Such models have been useful in providing a quantitative description of developmental landscapes that correspond to the spirit of Waddington's landscape, with different basins of attraction representing the valleys
of the differentiated states.

An important aspect of the reprogramming process is identifying the pathways through which a fully differentiated somatic cell is programmed back to pluripotency, and in particular, whether the path a cell takes in going from a somatic state to a pluripotent state is the same as the reverse pathway. Also of interest is characterising the possible intermediate states in the process. Recent experiments by Nagy and Nagy \cite{Nagy:10} have shed some light on the path the cell takes as it is reprogrammed back to a pluripotent state. They studied the reprogramming of differentiated secondary mouse fibroblast cells that were derived from induced pluripotent stem cells, and encoded the four Yamanaka factors under the control of doxycycline promoters. Thus expression of the four factors and induction of pluripotency in entire populations of the fibroblasts could be achieved by treating cultures with the drug doxycycline. They found that there were two distinct timescales in the reprogramming process, a Point of No-Return (PNR) time, below which, the cessation of the doxycycline input leaves the cell in the somatic state. The second characteristic time, called the Commitment to Pluripotent State (CPS) time, denotes the time beyond which application of doxycycline, commits the cell to the pluripotent cell fate. In between these two timescales, the PNR and the CPS, they found that the cell reached an undetermined state, which was neither somatic nor pluripotent, but rather signals the presence of a novel intermediate state in the reprogramming process. Cessation of the doxycycline input during this period results in neither return to somatic, nor progress to pluripotent states. They denoted this novel intermediate state as the ``Area 51" state. However, the characteristics of this state has not yet been determined.

The presence of an intermediate state in the reprogramming pathway promises to be an useful tool in understanding the mechanics of the uphill process. Further, a full understanding of the Area 51 state could lead to enhanced control over the reprogramming process, such as offering the possibility to create and maintain lineage-committed cells that have various applications. In this paper, we propose a theoretical framework that can lead to such intermediate states in the context of a gene-regulatory network. We propose that when modeling a gene network, an important physical factor that has so far not been taken into account is the effect of delays in the self-regulatory feedback mechanisms. The reprogramming of a somatic cell to pluripotency is a complex multistep reaction that involves both structural modifications to the chromatin network as well as changes in gene expression patterns~\cite{Carvalho:10}. These changes arise in response to the expression levels in the gene regulatory network, and are modeled by a self-regulating feedback loop. However since these changes occur in a finite time, the feedback loop should in fact depend on the state of the system at a previous instant of time, leading to delays. While delay differential equations have been used to study diverse systems~\cite{Mackey:77} such as modeling disease onset in physiological systems~\cite{Alexander:08}, and discrete time population models~\cite{Kuang:93}, we show that they may be critical in developing a mathematical framework for understanding the nature of the epigenetic landscape.

In this paper, we show the importance of time delays in the context of a gene-regulatory network. We model the regulatory network through the dynamics of a single differentiation regulator, denoted by $x$, that promotes its own synthesis through a feedback loop. While real life regulatory circuits in the cell depend on two or more differentiation regulators, the main aim of this paper is to show the effects of time delays in such circuits, and a single-variable genetic circuit offers a model system in which to study such effects. Such single variable circuits are similar to the models proposed for progesterone-induced Xenopus oocyte maturation~\cite{Ferrell:03,Ferrell:98,Ferrell:09,Ferrell:12}, and might also be applicable to scenarios where a single transcription factor such as $Myod$ has been shown to induce a change of cell fate from fibroblast to myoblast~\cite{Lassar:87}. We define the single-variable regulatory model in the next section, and discuss the results as a function of the parameters of the model. A discussion of the importance and applicability of the resulting phase diagram to systems of differentiating cells and its extension to more realistic gene regulatory networks are discussed in the final section.

\section{Model and Results}
Gene regulatory networks that control cell fate differentiation has been modeled by self-activating genes. While actual gene regulatory networks inside the cell may consist of multiple genes which have a complex interdependence on each other, one variable or two-variable gene networks provide an useful model to illustrate some of the basic principles of cell-fate determination.

\begin{figure}
\includegraphics[width=8cm]{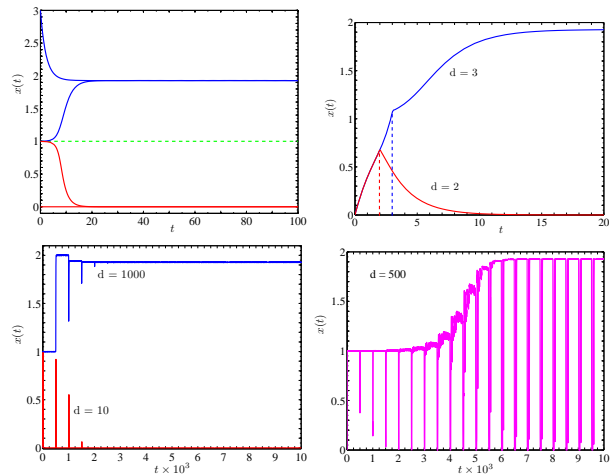}
\caption{Somatic ($x = 0$), induced pluripotent ($x \approx 2$), and Area $51$ cells in a single gene regulatory circuit. Panel (a) shows steady state values for Eq.~\ref{eq:one-ge-circuit} without drive or delay ($\alpha_0 = 0$, $d = 0$). Depending on the initial value $x(t=0)$, the somatic (red solid line) and the iPS cells (blue solid line) are stable. The unstable state $x = 1$ (green line) is also shown. Panel (b) shows corresponding steady states with a non-zero drive ($\alpha_0 = 0.5$), a decay constant $\beta = 0.5$, and the coefficient of self promotion $\alpha_1 = 1.0$. Depending on the duration $d = 2$ (red solid) the somatic, or $d = 3$ (blue solid) iPS cells are chosen. Panel (c) shows $x(t)$ vs.\ $t$ corresponding to Eq.~\ref{eq:one-ge-circuit} for a delay of $\tau = 500$ and for drive $d = 10$ (red line), and $d = 1000$ (blue line) indicating stability of somatic and iPS states. Panel (d) shows $x(t)$ vs.\ $t$ for $d = 500$ with sustained fluctuations between the iPS and somatic states.}\label{Figure1}
\end{figure}

We first introduce a single-variable model for cell differentiation, where a single regulator $x$ self-regulates its own synthesis, as proposed by Ferrell~\cite{Ferrell:12}. The equations governing the rate of change of expression of a single gene is given by
\begin{equation}
\frac{d x}{d t} =  \alpha_0 + \alpha_1 \frac{x^n}{S^{n} + x^n} - \beta x, \label{eq:ferrell}
\end{equation}
where the first term represents an external input $\alpha_0$ that is constantly applied. The second term represents a feedback dependant self-regulation, modelled by a Hill function of order $n$. The third term models degradation process through a mass action process with the degradation rate $\beta$. The right hand side of Eq.~\ref{eq:ferrell} can be integrated with respect to the variable $x$ to give an ``effective potential'' landscape having two stable minima corresponding to different levels of expression of the gene. This can be seen in Fig.~\ref{Figure1}(a). The two stable fixed points correspond to $x=\tilde{x}_{1}$, and $x=\tilde{x}_{2}$ respectively ($\tilde{x}_{1} = 0$, and $\tilde{x}_{2} \approx 2$ for $\alpha_{0}=0$) with an unstable extremum at $x=x^{*}$ ($x^{*} = 1$, for $\alpha_0 = 0$). In the absence of drive the final gene expression level is crucially dependent on its initial value $x(t=0)$. Therefore if $x(t=0) = [0, 1-\epsilon]$ the system approaches $x=\tilde{x}_{1}$, while if $x(t=0) = [1 + \epsilon, \infty]$, the fixed point $x=\tilde{x}_{2}$ is chosen. Furthermore, in this model beyond a critical value of the external input ($\alpha_0 > \alpha_c$), the minimum at $x=\tilde{x}_{1}$ becomes unstable and the long time steady state is always $x=\tilde{x}_{2}$. This is in line with Ferrell's idea that saddle-node bifurcations are inconsistent with Waddington's landscape picture as there are no alternative end point states. In his work Ferrell~\cite{Ferrell:12} further introduces a two variable gene regulatory circuit as a model mimicking lateral inhibition and demonstrates pitchfork bifurcation commensurate with Waddington's picture. A similar two variable model had been proposed around the same time by Wang \textit{et al.}\cite{Wang:11}.

Motivated by these gene regulatory network models that attempt at developing a quantitative picture of Waddington's landscape we propose a simple generic single-gene regulatory network model similar to Ferrell~\cite{Ferrell:12,Wang:11} incorporating time-dependent drive and delay. The rate of change of the gene regulator $x$ in this model is described by
\begin{equation}\label{eq:one-ge-circuit}
\frac{d x}{d t} = \alpha_0 \Theta \left[d - t \right] + \alpha_1 \frac{x^n(t-\tau)}{S^{n} + x^n(t-\tau)} - \beta x(t),
\end{equation}
where $\alpha_0$, $\alpha_1$ and $\beta$ have the same meanings as Eq.\ref{eq:ferrell}. However unlike that model both the chemical drive as well as the feedback are functions of time. The Heaviside function multiplying the $\alpha_0$ term represents the fact that the external input is applied for a finite time interval $d$, while the self-regulatory term is dependent on the state of the regulator $x$ at a previous instant of time $t-\tau$. The time delay in the self-regulation term in Eq.~\ref{eq:one-ge-circuit} can have several possible physical origins, including multi-step chemical reactions and cell shape changes. We have assumed no such delay in the degradation term as it does not have biochemical warrant at the same level as the self-regulation and it does not affect the general results in our model.

We numerically integrated Eq.~\ref{eq:one-ge-circuit} for different values of the delay time $\tau$ and drive $d$. Figure \ref{Figure1}(b) represents the results of the single gene regulatory circuit without delay and with a chemical drive acting for a finite interval $d$ on an initial state $x = 0$. The self-promotion rate coefficient is $\alpha_1 = 1$ and the decay constant $\beta = 0.5$. Further, the amplitude of the chemical drive is parameterised by $\alpha_0 = 0.5$. We find that for a value of $\alpha_0 < \alpha_c$ and the duration of the drive $d$ less than a critical value $d_c (\approx 2$), the long time steady state is $x=0$. If however the drive is applied for a duration longer than $d_c$, starting from a state $x(t=0) = 0$ the system transitions to the other minimum $x \approx 2$. Identifying the $x=0$ state as a somatic and $x \approx 2$ as the pluripotent state, the above process describes inducing pluripotency via a chemical drive.

Figure~\ref{Figure1}(c) shows the variation of $x(t)$ vs $t$ starting from the somatic state $x=0$ for $d=10$, and $d=1000$, and a time delay $\tau = 500$ for the same set of parameters $\alpha_0$, $\alpha_1$ and $\beta$. As seen in the figure for $d=10$, the system relaxes back to the $x=0$ steady state, while for $d=1000$ the pluripotent state $x \approx 2$ is chosen. Sharp spikes showing attempted transitions between the two states are also seen. In the intermediate regime when the drive $d$ is of the same order of magnitude as the delay $\tau$, the trajectory of $x(t)$ shows sustained oscillations. This is shown in Fig.~\ref{Figure1}(d). We interpret such sustained oscillations as the cells which are caught in a limbo between the pluripotent and the somatic states and conjecture that these states are possibly the ones seen in the experiments by Nagy et al.~\cite{Nagy:10} termed ``Area 51''. The chemical drive $\alpha_0$ is then interpreted as the doxycycline input to somatic cells having a non-zero value, corresponding to a finite rate of basal synthesis, which is switched off ($\alpha_0 = 0$) beyond the input time.
\begin{figure}
\includegraphics[width=8cm]{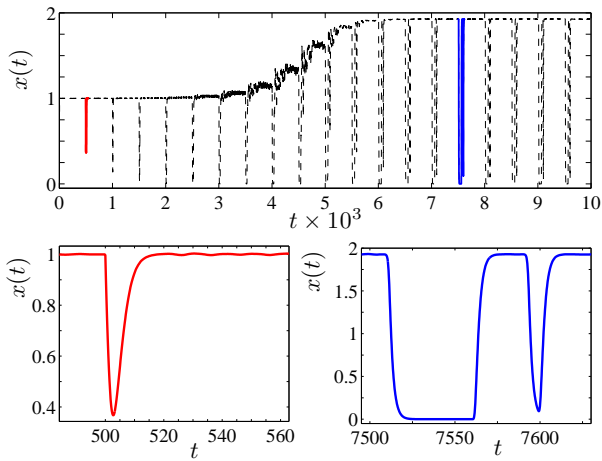}
\caption{Fluctuations in the ``Area 51'' region as a combined result of time-dependent drive and delay. The top panel shows sustained oscillations for the parameters of Fig.~\ref{Figure1}(d). The bottom panels indicate the oscillations in the transient ($500 \leq t \leq 540$) and sustained oscillatory ($7500 \leq t \leq 7650$) regions.}\label{Fig2illustrate}
\end{figure}

The oscillations seen in some solutions of Eq.~\ref{eq:one-ge-circuit} is an inherent feature of delay differential equations~\cite{Mackey:77}. These oscillations as shown in Fig.~\ref{Figure1}(d) are investigated in greater detail in figure~\ref{Fig2illustrate}. It is possible to analyse the time of occurence of these sharp spikes. If the drive duration is smaller than the delay time, \textit{i.e.} $d < \tau$, $x$ initially increases from its zero value as a function of time. Once the drive is withdrawn the dynamics of the system is completely dominated by the degradation term and as a result $x$ decreases. This behavior continues till $t = \tau$ when the self-regulation term promoting gene activity becomes non-zero, and as a result $x$ increases monotonically till a time $d + \tau$. At this time the self-regulatory term picks up the values of $x$ from the earlier cycle which was dominated by degradation kinetics. This can be generalised to state that the downward spikes occur at $t_p = d + p \tau$, while the upturns occur at $t = q \tau$. The slope of the first downturn is completely dictated by $\beta$ while the upturn slope turns out to be a
nonlinear function of $\alpha_1$ and $\beta$. For the situation in which $d > \tau$ the first upward turn occurs at $t = \tau$ followed by a downturn upon reduction of the drive at $t = d + \tau$. Following this oscillations are repeated at $t=t_p$ as discussed above. The preceding analysis is strictly valid in the initial time regime, where the spikes occur singly, as shown in Fig.~\ref{Fig2illustrate}(b). At later times, the single spikes give way to a double spike, with two spikes occurring in quick succession, as shown in Fig.~\ref{Fig2illustrate}(c). A complete description of the behaviour of the oscillations in this later time regime requires a full non-linear analysis of the original equation.

\begin{figure}
\includegraphics[width=8cm]{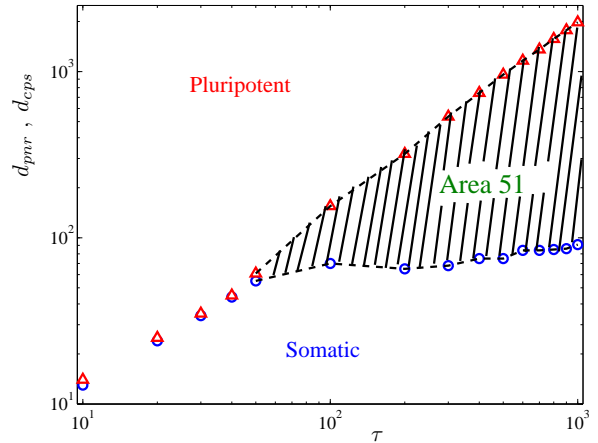}
\caption{Phase diagram showing regions where somatic and pluripotent states are stable as a function of the delay time $\tau$. The phase boundaries indicating point of no return (blue open circles), $d_{PNR}$, and those committed to the pluripotent state (red triangles), $d_{CPS}$ are indicated. The region between the two states mark the region when the cell fate attains neither fixed point, but oscillates indefinitely, termed ``Area 51''\cite{Nagy:10}.}\label{Figure2}
\end{figure}

The two critical time scales alluded to earlier, the ``point of no-return'', and ``commitment to pluripotent state''is seen in Figure \ref{Figure1}(c) and (d) respectively. These indicate threshold values such that for $d < d_{PNR}$ the system would return to their somatic state, while for $d > d_{PNR}$ the cell fate is changed. The second threshold corresponds to the drive being on for a duration $d > d_{CPS}$ which results in a final pluripotent cellular state. The intermediate region of drives $d_{PNR} < d < d_{CPS}$ defines the ``Area 51'' region. Taking cue from our numerical results discussed above we draw a phase diagram showing the domain of ``Area 51'' as functions of $d$ and $\tau$ in a single gene regulatory circuit incorporating time dependent drive and delay dynamics.

Figure~\ref{Figure2} demonstrates the variation of the two thresholds $d_{PNR}$ and $d_{CPS}$ as a function of the delay $\tau$. For $0 \leq \tau \leq 50$, the two threshold values are almost the same, \textit{i.e.} $d_{PNR} \approx d_{CPS}$. In this regime the system transitions from the somatic state to the induced pluripotent state once the duration of the drive is greater than $d_{PNR}$. However for larger values of $\tau$ the two threshold values are different exposing an intermediate regime marked by sustained oscillations. As seen from the graph $d_{CPS}$ monotonically increases with delay $\tau$ while some fluctuations in $d_{PNR}$ is observed. With increasing $\tau$ the ``Area 51'' region widens as can be seen in Figure~\ref{Figure2}.



\section{Discussion}

We have illustrated the importance of time delays in feedback circuits in the context of a simple gene regulatory network in which the state of differentiation is regulated by a single differential regulator. The energy landscape of the model, in the absence of delays, has two minimas, denoting the pluripotent and differentiated states. Introducing a delayed self-regulation term changes the landscape such that there is now a region in phase space, in which the system has a long-lived oscillatory nature. We propose that such oscillatory states may underlie the existence of novel intermediate states observed in the reprogramming of mouse somatic cells, and denoted by ``Area 51''. Further experiments with fast decaying reporters which are proxies for pluripotency or somatic cell markers would be needed to validate our hypothesis of a fluctuating intermediate state.


In order to model more realistic differentiation events, one would need to study higher dimensional systems where the number of differential regulators is more than one. Two variable gene-regulatory models~\cite{Wang:11} offer a straight forward generalisation of these ideas to mimic realistic cell differentiation scenarios. For a full description of the dynamics of the reprogrammed cell due to the four Yamanaka factors, one needs to study the effect of delays in a four variable model, and map out the effect of the interplay of these four variables on the intermediate state.

The switch from the somatic state to the pluripotent state is accompanied by various changes inside the cell, including changes in the chromatin structure, loss of somatic cell specific markers, and reactivation of endogenous genes essential for pluripotency and self-renewal, among others. Recent experiments suggest that the various changes associated with pluripotency occur in a well-defined sequential manner. For instance, the pluripotency marker of mouse pluripotent cells, SSEA-1 appears to be expressed in the very early stages of pluripotency ~\cite{Jaenisch:08,Stadtfeld:08}, while the reactivation of endogenous genes such as Oct4, Nanog and Sox2 occurs late in the reprogramming process. It is probable that the rapid fluctuations predicted by the delayed-self-regulation model proposed here arise only in the context of one or a few of these pluripotency markers, instead of the full state of the cell switching from somatic to pluripotent. Thus experiments designed to validate this hypothesis of a
fluctuating intermediate state need to identify the probable candidates for such switching.

Another area of interest in the context of induced pluripotent cells is whether there is an inherent asymmetry to the landscape. Nagy \textit{et al.} does not comment whether the ``Area 51'' is encountered if we perform the reverse experiment, \textit{i.e.} start from the pluripotent state and induce differentiation by keeping the cells in a chemical environment for different durations. Further experiments are needed to map out the landscape as a pluripotent cell divides under the influence of a time-dependent stimuli. Such experiments would then provide an additional input to the model to facilitate understanding of the full epigenetic landscape.

The concept of time delays, possibly induced by remodelling of cellular architecture, is an important one in the differentiation context, as reorganization events inside the cell that accompany a change in cell state take place over a time scale of days~\cite{Foster:11}. Thus when modelling the epigenetic landscape through dynamical equations, one must consider the effect of delays on differentiation pathways. Similar oscillatory behavior has also been observed in other related biological systems, such as the Epithelial to Mesenchymal transition in early embryonic development and cancer metastasis~\cite{ConacciSorrel:03,Vuoriluoto:11}. In both these situations the oscillations arise from time dependent remodelling of the cytoskeleton. Thus the concept of delays may be important in other biological contexts too and should prove a useful tool in the design of predictive experiments.


\section{Acknowledgements}

MM $\&$ BC acknowledges financial support from Institute of Advanced Studies Durham University, the Department of Mathematical Sciences, and the Biophysical Sciences Institute, via the IAS-BSI COFUND fellowship. BC acknowledges hospitality of the Isaac Newton Institute, Cambridge University.


\begin{thebibliography}{99}

\bibitem{Solter:84} J. McGrath and D. Solter, Science, {\bf 226}, 1317 (1984).

\bibitem{Gurdon:62} J. B. Gurdon, J. Embryol. Exp. Morph., {\bf 10}, 622 (2006).

\bibitem{Gurdon:06} J. B. Gurdon, Annu. Rev. Cell  Dev. Biol., {\bf 22} 1 (2006).

\bibitem{Yamanaka:06} K. Takahashi and S. Yamanaka, Cell, {\bf 126}, 663 (2006).

\bibitem{Yamanaka:07} K. Okita, T. Ichisaka and S. Yamanaka, Nature, {\bf 448}, 313 (2007).

\bibitem{Jaenisch:07} N. Maherali, R. Sridharan, W. Xie, J. Utikal, S. Eminli, K. Arnold, M. Statfeldt, R. Yachechko, J. Tchieu, R. Jaenisch, Cell Stem Cell, {\bf 1}, 55 (2007).

\bibitem{Hochedlinger:08} N. Maherali, T. Ahfeldt, A. Rigamonti, J. Utikal, C. Cowan and K. Hochedlinger, Cell Stem Cell, {\bf 3}, 340 (2008).

\bibitem{Wernig:07} M. Wernig, A. Meissner, R. Foreman, T. Brambrink, M. Ku, K. Hochedlinger, B.E. Bernstein, R. Jaenisch, Nature, {\bf 448}, 318 (2007).

\bibitem{Carvalho:10} T. K. Kelly, D. D. De Carvalho, and P.A. Jones, Nature Biotechnology, {\bf 28}, 1069 (2010): D. D. De Carvalho, J. S. You, and P. A. Jones. Trends in Cell Biology {\bf 20}, 609 (2010).

\bibitem{Waddington:57} C. H. Waddington, \textit{The Strategy of Genes}, Allen $\&$ Unwin, London (1957).

\bibitem{Wang:11} J. Wang, K. Zhang, X. Liu, E. Wang, Proc. Natl. Acad. Sci., {\bf 108}, 8257 (2011).

\bibitem{Ferrell:12} J. E. Ferrell Jr., Curr. Biol., {\bf 22}, R458 (2012); J. E. Ferrell Jr., {\textit et al.} FEBS Lett., {\bf 583}, 3999 (2009); J. E. Ferrell Jr, W. Xiong, Chaos, {\bf 11}, 227 (2001);

\bibitem{Nagy:10} A. Nagy, and K. Nagy, Nature Methods, {\bf 7}, 22 (2010).

\bibitem{Yu:07} J. Yu, M.A. Vodyanik, K. Smuga-Otto, J. Antosiewicz-Bourget, J.L. frane, S. Tian, J. Nie, G.A. Jonsdottir, V. Ruotti and R. Stewart, Science, {\bf 318}, 1917 (2007).

\bibitem{Thomson:09} J. Yu, K. Hu, K. Smuga-Otto, S. Tian, R. Stewart, I.I. Sluvkin and J.A. Thomson, Science, {\bf 324}, 797 (2009).

\bibitem{Huang:10} J.X. Zhou, S. Huang, Trends Genet, {\bf 27}, 55 (2010).

\bibitem{Enver:07} S. Huang, Y.P. Guo, G. May, T. Enver, Dev. Biol. {\bf 305}, 695 (2007).

\bibitem{Enver:09} T. Graf, T. Enver, Nature, {\bf 462}, 587 (2009).

\bibitem{Lassar:87} R.L. Davis, H. Weintraub and A.B. Lassar, Cell, {\bf 51}, 987 (1987).

\bibitem{Ferrell:03} W. Xiong and J.E. Ferrell, Jr., Nature, {\bf 426}, 460 (2003).

\bibitem{Ferrell:98} J.E. Ferrell, Jr. and E.M. Machleder, Science, {\bf 280}, 895 (1998).

\bibitem{Ferrell:09}  J.E. Ferrell, Jr., J.R. Pomerening, S.Y. Kim, N.B. Trunnell, W. Xiong, C.Y. Huang and E.M. Machleder, FEBS Lett. {\bf 583}, 3999 (2009).

\bibitem{Jaenisch:08} T. Brambrink, R. Foreman, G.G. Welstead, C.J. Lengner, M. Wernig, H. Suh, R. Jaenisch, Cell Stem Cell, {\bf 2}, 151 (2008).

\bibitem{Stadtfeld:08} M. Stadtfeld, N. Maherali, D.T. Breault, K. Hochedlinger, Cell Stem Cell, {\bf 2}, 230 (2008).

\bibitem{Rizzino:10} J.L. Cox and A. Rizzino, Expt. Biol. and Med., {\bf 235}, 148 (2010).

\bibitem{Hochedlinger:10} M. Stadtfeld and K. Hochedlinger, Genes Dev., {\bf 24}, 2239 (2010).

\bibitem{Mackey:77} M.C. Mackey and L. Glass, Science, {\bf 197}, 287 (1977).

\bibitem{Alexander:08} M. E. Alexander, S. M. Moghadas, Gergely R\"{o}st, Jianhong Wu, Bull. Math. Biol., {\bf 70}, 382 (2008).

\bibitem{Kuang:93} \textit{Delay Differential Equations: With Applications in Population Dynamics}, ed. Y. Kuang, Academic Press, Inc. (1993).
    
\bibitem{Foster:11} C. R. Foster, J. L. Robson, W. J. Simon, J. Twigg, D. Cruikshank, R. G. Wilson, and Christopher J. Hutchison, Nucleus, {\bf 2}, 5 (2011).

\bibitem{ConacciSorrel:03} M. E. Conacci-Sorrell, I. Simcha, T. Ben-Yedidia, P. Savagner, and A. Ben-Ze\'{e}v, J. Cell Biol. {\bf 163}, 847 (2003): D. Chen, W. Xu, E. Bales, C. Colmenares, M. E. Conacci-Sorrell, D. Eling, S. Ishii, E. Stavneze, J. Campisi, D. Fisher, A. Ben-Ze\'{e}v, and E. Medrano, Cancer Res. {\bf 63}, 6626 (2003).

\bibitem{Vuoriluoto:11} K. Vuoriluoto, H. Haugen, S. Kiviluoto, J. P. Mpindi, J. Nevo, C. Gjerdrum, C. Tiron C, J. B. Lorens, and J. Ivaska, Oncogene. {\bf 30}, 1436 (2010).

\end{thebibliography}
\end{document}